\documentclass[prd,twocolumn,nofootinbib,preprintnumbers,floatfix]{revtex4}  % for review and submission
\usepackage[plainpages=false, colorlinks=true, anchorcolor=blue, linkcolor=blue, citecolor=blue, bookmarks=false]{hyperref}
\usepackage{graphicx}  % needed for figures
\usepackage{dcolumn}   % needed for some tables
\usepackage{bm}        % for math
\usepackage{amssymb}   % for math
\usepackage{natbib}
\usepackage{amsmath}   %%%%   added by me
\usepackage{longtable}  %%% added by kamal
\usepackage{float}    %%% added by kamal
\usepackage{tasks}
\begin{document}
\newcommand{\apjl}{Astrophys. J. Lett.}
\newcommand{\apjs}{Astrophys. J. Suppl. Ser.}
\newcommand{\aap}{Astron. \& Astrophys.}
\newcommand{\rthis}[1]{\textcolor{black}{#1}}
\newcommand{\aj}{Astron. J.}
\newcommand{\pasp}{PASP}
\newcommand{\araa}{Ann. Rev. Astron. Astrophys. } %ARA$\&$A}
\newcommand{\aapr}{Astronomy and Astrophysics Review}
\newcommand{\ssr}{Space Science Reviews}
\newcommand{\mnras}{Mon. Not. R. Astron. Soc.}
\newcommand{\apss} {Astrophys. and Space Science}
\newcommand{\jcap}{JCAP}
\newcommand{\na}{New Astronomy}
\newcommand{\pasj}{PASJ}
\newcommand{\pasa}{Pub. Astro. Soc. Aust.}
\newcommand{\physrep}{Physics Reports}

% The following information is for internal review, please remove them for submission

% the following line is for submission, including submission to the arXiv!!
%\hspace{5.2in} \mbox{Fermilab-Pub-04/xxx-E}

\title{A model-independent test of  the  evolution of gas depletion factor for SPT-SZ and Planck ESZ clusters}

\author{Kamal 
\surname{Bora}}
 \altaffiliation{E-mail: ph18resch11003@iith.ac.in}
 
\author{Shantanu \surname{Desai}}%
\altaffiliation{E-mail: shntn05@gmail.com}

%\date{\today}

\begin{abstract}
The gas mass fraction in galaxy clusters has been widely used to determine cosmological parameters. This method assumes that the ratio of the cluster gas mass fraction to the cosmic baryon fraction ($\gamma(z)$)   is constant as a function of redshift. In this work, we look for a time evolution of $\gamma(z)$ at $R_{500}$  by using both the SPT-SZ and Planck Early SZ (ESZ) cluster data, in a model-independent fashion without any explicit dependence on the underlying cosmology. For this calculation, we use a non-parametric functional form for the Hubble parameter obtained from Gaussian Process regression using cosmic chronometers. We parameterize $\gamma(z)$ as:  $\gamma(z)= \gamma_0(1+\gamma_1 z)$ to constrain the redshift evolution. We find contradictory  results between both the samples. For SPT-SZ,  $\gamma (z)$ decreases as a function of redshift (at more than 5$\sigma$), whereas a positive trend with redshift is found for Planck ESZ data (at more than 4$\sigma$). 
We however find that the $\gamma_1$ values for a subset of SPT-SZ and Planck ESZ clusters  between the same redshift interval agree to within $1\sigma$. When we allow for a dependence on the halo mass in the evolution of the gas depletion factor, the $4-5\sigma$ discrepancy  reduces to $2\sigma$.

\end{abstract}

\affiliation{
 Department of Physics, Indian Institute of Technology, Hyderabad, Kandi, Telangana-502285, India} 
\maketitle
%\pacs{}

%\section{\label{sec:level1}First-level heading}
% sections are not used for PRL papers

\section{Introduction}

Galaxy clusters are the most massive virialized objects in the universe and provide wonderful laboratories for studying a wide range of topics from galaxy evolution to cosmology~\cite{Vikhlininrev,allen11} to fundamental Physics~\cite{Desai}.  They  are also key in pinning down the dark energy equation of state, as they probe both the geometry of the Universe and the growth of structure~\cite{Huterer,Weinberg}. Galaxy clusters are   therefore, one of the flagship probes for Stage IV dark energy experiments~\cite{Albrecht}. In the past two decades, a large number of galaxy clusters have been discovered upto very high redshifts, thanks to myriad multi-wavelength surveys, which have greatly facilitated these  science goals.

Since galaxy clusters are the largest gravitationally bound objects, their matter content present in the intracluster medium has provided a powerful probe to constrain the cosmological parameters. The X-ray emitting gas in the intra-cluster medium  dominates the baryon budget in clusters. The idea of using the gas mass fraction as a cosmological tool was first introduced by White et al~\cite{white93}.  The gas mass fraction, which is defined as $f_{gas} \equiv M_{gas} / M_{tot}$ is expected to  match closely  the cosmic baryon fraction, viz. $\Omega_b / \Omega_m$, where $\Omega_b$ and  $\Omega_m$ are the cosmic baryon density and total matter density respectively, in units of critical density ($\rho_c$). 
There have been a large number of studies, constraining the matter density ($\Omega_m$), dark energy equation of state parameter ($\omega$), and the curvature density ($\Omega_k$)  using the $f_{gas}$ measurements~\cite{Sasaki96,Pen,lima03,allen02,allen04,allen08,ettori03,ettori09,mantz14}. The $f_{gas}$ test has also been used to test alternatives to $\Lambda$CDM such as $R_h=ct$ universe~\cite{Melia16}. The main assumption in these studies is that  the gas fraction doesn't evolve with redshift for hot, massive and relatively relaxed clusters.
However, the  evolution  of $f_{gas}$  as a function of redshift needs to be determined in a model-independent method, in order to probe the robustness of these results. 
This evolution is usually parametrized  using the   gas depletion factor $\gamma$, which is the ratio by which the baryonic gas fraction in galaxy cluster gets depleted by the cosmic mean baryon fraction, and is given by: $\gamma  =  f_{gas} (\Omega_b / \Omega_m)^{-1}$~\cite{allen08}. So the study of gas depletion factor and it's evolution with redshift  play a pivotal  role in the  robustness of  the $f_{gas}$ test. As argued in ~\cite{holanda1706}, the results for  the gas depletion factor obtained using only  the data could  then be used as a prior for any cosmology studies with $f_{gas}$.

Therefore, a large number of studies have been undertaken using both simulations and data to test the variation  of the gas depletion factor $\gamma(z)$ as a function of redshift. This evolution  is usually  modeled using the following function.
\begin{equation}
    \gamma(z)= \gamma_0(1+\gamma_1 z)
    \label{eq:gammavariation}
\end{equation}
 where $\gamma_0$ is the normalization factor and $\gamma_1$ represents the evolution of gas depletion factor with redshift.
%Now a days, improved hydrodynamical simulations of galaxy cluster formation showed a negligible evolution of gas depletion factor with redshift. 

Battaglia et al~\cite{battaglia13} studied the measurement biases of $f_{gas}$ using   hydrodynamical simulations including different kinds of processes like star formation, AGN feedback and radiative physics. They found a constant gas mass fraction at $R_{500}.$\footnote{$R_{500}$ is the radius at which the mass density is 500 times the critical density for Einstein-DeSitter universe~\cite{White01}.} %Their non-radiative simulations exclude the radiative cooling flow in the cluster, hence not able to resolve much about the formation and evolution of the galaxies presents within the clusters~\cite{evrard90}. However, on the other hand, radiative simulations incorporate the cooling effect of the gas, a part of which leads to the star formation and rest of the reheated by the star itself. Although reheating could also be done by the AGN feedback~\cite{kay04}. 
Thereafter, Planelles et al~\cite{planelles13} studied the baryon fraction within $R_{2500}$, $R_{500}$, and $R_{200}$ by using three different suites of  simulations of  galaxy clusters in the redshift range $0 \leqslant z \leqslant 1$ for different physical processes in clusters. The first set of simulations (NR) involved non-radiative  hydrodynamical simulations. The second suite (CSF)  studied  the effects of cooling,  supernovae feedback, and star formation. Finally the third set of simulations (AGN) included the same physics as CSF, along with   AGN feedback. Furthermore, they discussed the baryon  and gas depletion factor and its dependence on the  cluster radius, redshift and baryonic physics. They concluded that the evolution of the depletion factor is negligible with redshift (for $z < 1$) regardless of the cluster radius and baryonic physics. This aforementioned work obtained: $\gamma_0 = 0.79\pm0.07$ and $\gamma_1 = 0.07\pm0.12$ at $R_{2500}$, $\gamma_0 = 0.85\pm0.03$ and $\gamma_1 = 0.02\pm0.05$ at $R_{500}$, $\gamma_0 = 0.86\pm0.02$ and $\gamma_1 = 0.00\pm0.04$ at $R_{200}$~\cite{planelles13}. We note that all these aforementioned simulations assumed $\Lambda$CDM as the base cosmological model.

Holanda et al.~\cite{holanda1706} made the first attempt to study the gas depletion factor using only  observations. They carried out an analysis using 40 $f_{gas}$ measurements~\cite{mantz14} observed by the Chandra X-ray telescope spanning the redshift range $0.078 \leqslant z \leqslant 1.063$ characterized as massive, morphologically relaxed systems and with $kT \geqslant 5$~keV. In their work, $f_{gas}$ measurements were obtained from inside a shell with radii between (0.8 - 1.2) $R_{2500}$. Their analysis assumed the validity of cosmic distance duality relation (CDDR)~\cite{ellis07}, and  the use of different distance indicators, e.g. Type Ia SNe (from Union 2.1 and JLA compilation) and $\Lambda$CDM model.  They  obtained  $\gamma_0 = 0.86\pm0.04$ and  $\gamma_1 = -0.04\pm0.12$~\cite{holanda1706}.  Therefore, their analysis showed no evolution of  the gas depletion factor with redshift. 

Then, in a followup work, ~\citet{holanda1711} used 38 Chandra X-ray  $f_{gas}$ measurements  from~\cite{laRoque06} in the redshift range $0.14 \leqslant z \leqslant 0.89$, along with angular diameter distances from X-ray/SZ measurements. Unlike ~\cite{holanda1706}, they did not use CDDR to derive the angular diameter distance. They reported $\gamma_0$ =  $0.76\pm0.14$ and $\gamma_1$ = $-0.42_{-0.40}^{+0.42}$ at $R_{2500}$, which also indicates no time evolution for the  gas depletion factor.

Most recently, ~\citet{zheng19} (Z19, hereafter) did a similar study using data for 182 clusters with  the SZ~\cite{SZ,birki,carl} selected sample from the  Atacama Cosmology Telescope Polarization experiment (ACTPol) spanning the redshift range $0.1 \leqslant z \leqslant 1.4$~\cite{actpol}.  The main difference between Z19 and the previous works~\cite{holanda1706,holanda1711} is that Z19 considered the data at $R_{500}$, whereas the latter considered $R_{2500}$.  
However,  Z19
did not use direct $f_{gas}$ measurements, but instead
resorted to  a  semi-empirical relation between  the gas mass fraction   and the total cluster mass~\cite{vikhlinin09}. To estimate the angular diameter distance (needed for the estimation of the  gas depletion factor), a model-independent approach using Gaussian Processes regression was used. They reported that the  gas depletion factor decreases as a function of redshift. 
Similar to Z19, we study  the evolution of  the gas depletion factor for two different SZ selected cluster samples. The first set comprises  a sample 94 South Pole Telescope (SPT) selected clusters in the redshift range $0.278 \leqslant z \leqslant 1.32$, and the second set uses 120 Planck Early SZ (ESZ) data covering the redshift range $0.059 \leqslant z \leqslant 0.546$. 

This paper is structured as follows. In section~\ref{sec:sample}, we briefly describe both our cluster samples, and their $f_{gas}$ measurements. In section~\ref{sec:method}, we describe the procedure used to determine the evolution of  gas mass fraction. Our analysis along with results are discussed in section~\ref{sec:analysis}. A discussion of our results can be found in section~\ref{sec:dicussions}. Finally, we conclude in section~\ref{sec:conclusion}.

\section{Galaxy Cluster Sample and their \boldmath$f_{gas}$ measurements}
\label{sec:sample}

\subsection{For SPT-SZ Sample}
The SPT-SZ cluster sample used for this analysis, consists of  91 SPT clusters from~\cite{chiu18} (C18, hereafter) and three additional SPT-SZ clusters: namely SPT-CLJ 0205-5829, SPT-CLJ 0615-5746 and SPT-CLJ2040-5726 from~\cite{chiu16}, giving a total of 94 clusters, with a mass threshold of   $M_{500} \geqslant 2.5 \times 10^{14}$ $M_\odot$, and  in the redshift range of $0.278 \leqslant z \leqslant 1.32$. These clusters  were detected in the  2,500 deg$^2$ South Pole Telescope (SPT) SZ survey~\cite{carlstrom11}. In this work, we used redshifts for each of these clusters from ~\citet{bocquet19}.  The SPT is a 10~m telescope at the South Pole, that  has imaged the sky at three different frequencies, viz. 95 GHz, 150 GHz, and 220 GHz~\cite{carlstrom11}. The SPT collaboration  carried out a 2500 square degree survey between 2007 and 2011 to detect galaxy clusters using the SZ effect.
This SPT-SZ survey detected 516 galaxy clusters with a mass threshold of $3 \times 10^{14} M_{\odot}$ upto redshift of  1.8~\cite{bleem15,bocquet19}. Detailed properties of the SPT  clusters  are discussed in~\cite{bleem15}. Their redshifts  have been obtained using survey data as well as   pointed spectroscopic and photometric observations~\cite{song12,ruel14,bayliss17,desai12,saro15}. 
%Chiu et al.~\cite{} used the SZE scaling relation from ~\cite{dehaan16} to estimate the total mass of the clusters, $M_{500}$ (for details, see~\cite{bocquet15}) while $M_{gas}$(denoted by $M_{ICM}$ in~\cite{chiu18}) is obtained by integrating the best-fit modified $\beta$-model~\cite{vikhlinin06} within $R_{500}$(for more details, see~\cite{mcdonald13}). From there, we calculate the gas mass fraction $f_{gas}$, as the ratio of $M_{gas}$ to $M_{500}$.
The first step in estimating the gas mass fraction consists of  determining  the total mass of the cluster at $R_{500}$, known as  $M_{500}$. This was estimated from the SZ
detection significance or signal to noise ratio, using SZ observable to mass scaling relations discussed  in ~\cite{bocquet15,dehaan16}. These scaling relations have  been self-consistently calibrated  using
a combination of X-ray, weak lensing, and number count distributions~\cite{chiu18}.
Here, we briefly describe the procedure used to derive the $f_{gas}$ measurements below. More details can be found in the aforementioned works.

 C18 estimated the total mass (denoted by $M_{500}$) and gas mass (which they refer to as intra-cluster medium mass, denoted by $M_{ICM}$) from the analysis of Chandra X-ray observations. All the 91 clusters have been imaged with Chandra, either through the Chandra X-ray Visionary Project led by the SPT collaboration as well as  other proposals led by non-SPT members.

To estimate $M_{ICM}$, a modified-$\beta$~\cite{vikhlinin06} model is used  to fit the gas density, which was estimated from the observed surface brightness  profile in the energy range 0.7-2.0 keV out to 1.5 $R_{500}$. This surface profile was then fit to the  modified $\beta$-model~\cite{vikhlinin06}. More details of this analysis can be found in ~\cite{mcdonald13}. 
We then calculate the gas mass fraction $f_{gas}$, as the ratio of $M_{ICM}$ to $M_{500}$. This estimated gas mass fraction as a function of redshift is shown in Fig.~\ref{fig:f1}. We note that C18  also investigated the trends of gas mass  fraction as a function of redshift. However the parametric form which they have considered (Eq. 6 in C18) is different than that considered in this work (cf. Eq.~\ref{eq:gamma_obs}), and their analysis also includes a dependence  on $M_{500}$. A comparison between our results will be discussed in Sec.~\ref{sec:dicussions}. 

Another thing to point out is that for calculating this gas mass fraction, $M_{500}$ was estimated using the observable SZ to mass scaling relation derived in a previous SPT work~\cite{dehaan16}. However, this scaling relation can change when using the cosmology priors from the Planck CMB anisotropy observations~\cite{bocquet15}. If this cosmology is adopted, $M_{500}$ would be larger by about 22\%, which in turn would also effect the baryon fraction numbers computed in C18.  However, the Planck Cosmology prior is disfavored, since the weak lensing and dynamical masses are in agreement with the standard SPT cluster analysis~\cite{chiu18}. In this work, we therefore do not consider the impact from the Planck cosmology prior.

\subsection{For Planck ESZ Sample}
In this work, we use the gas mass fraction measurements of  120 Planck Early SZ (ESZ) clusters~\cite{Planck11}  spanning the redshift range $0.059 \leqslant z \leqslant 0.546$~\cite{lovisari20}. All these clusters were also  imaged  with XMM-Newton up to $R_{500}$~\cite{lovisari17}. The only difference compared to this dataset  is that we used updated redshifts from Planck SZ2 catalog~\cite{PlanckSZ2}, instead of the ESZ redshifts used in ~\cite{lovisari17}.  These redshifts have also been obtained  using pointed optical follow-ups with various telescopes (eg.~\cite{Liu15})
and are still been continuously refined (eg.~\cite{Klein,Zaznobin}). Therefore, the redshifts of some of these clusters have further been updated after the publication of the Planck SZ2 catalog. The redshifts used for this analysis can be found in Table~\ref{tab:updated z}.  
%Asof sosumfts ing spherical symmetry, Lovisari etal.~\cite{lovisari20} derived the total mass of the cluster at $R_{500}$ by using hydrostatic equilibrium condition whereas $M_{gas}$ is obtained by integrating the density profile(~\cite{lovisari17}) within $R_{500}$. The ratio of mass of the gas to the total cluster mass i.e. $M_{gas} / M_{500}$ gives the gas mass fraction $f_{gas}$ at $R_{500}$.

Lovisari et al.~\cite{lovisari20}  derived the total mass of the cluster at $R_{500}$ from the equation of hydrostatic equilibrium  and assuming spherical symmetry: 
\begin{equation}
	\label{eq:hydrostatic}
    M(r<r_{500}) = -\frac{kT_{gas}}{G\mu m_p{}{}} {\left[\frac{d\log\rho_{gas}}{d\log r}  + \frac{d\log T_{gas}}{d\log r}\right]}
\end{equation}
where $T_{gas}(r)$ and $\rho_{gas}(r)$ denote the temperature and density profiles (see~\cite{lovisari20,lovisari17}), $k$ is the Boltzmann's constant and $\mu m_p$ represents the mean molecular weight of the intra-cluster gas.

\begin{longtable}[t]{|l|c|c|}  
%	\centering
\caption{ Redshifts of Planck-ESZ clusters used for this analysis.} \label{tab:updated z} \\
	\hline
	 \textbf{Cluster} & \textbf{Redshift} & \textbf{Reference} \\
	\hline
PSZ2 G000.40-41.86 & 0.1651 & ~\cite{PlanckSZ2}\\
PSZ2 G002.77-56.16 & 0.1411 & ~\cite{PlanckSZ2}\\ 
PSZ2 G003.93-59.41 & 0.151 & ~\cite{PlanckSZ2}\\
PSZ2 G006.68-35.55 & 0.0894 & ~\cite{PlanckSZ2}\\
PSZ2 G006.76+30.45 & 0.203 & ~\cite{PlanckSZ2}\\
PSZ2 G008.47-56.34 & 0.1486 & ~\cite{PlanckSZ2}\\
PSZ2 G008.94-81.22 & 0.3066 & ~\cite{PlanckSZ2}\\
PSZ2 G021.10+33.24 & 0.1514 & ~\cite{PlanckSZ2}\\
PSZ2 G036.73+14.93 & 0.1525 & ~\cite{PlanckSZ2}\\
PSZ2 G039.85-39.96 & 0.176 & ~\cite{PlanckSZ2}\\
PSZ2 G042.81+56.61 & 0.0731 & ~\cite{rines16}\\
PSZ2 G046.10+27.18 & 0.389 & ~\cite{PlanckSZ2}\\
PSZ2 G046.47-49.44 & 0.0846 & ~\cite{PlanckSZ2}\\
PSZ2 G049.22+30.87 & 0.1604 & ~\cite{rines16}\\
PSZ2 G049.32+44.37 & 0.096 & ~\cite{rines16}\\
PSZ2 G049.69-49.46 & 0.098 & ~\cite{PlanckSZ2}\\
PSZ2 G053.53+59.52 & 0.1132 & ~\cite{rines16}\\
PSZ2 G055.59+31.85 & 0.2242 & ~\cite{rines16}\\
PSZ2 G055.95-34.89 & 0.2306 & ~\cite{rykoff16}\\
PSZ2 G056.77+36.32 & 0.0997 & ~\cite{rines16}\\
PSZ2 G056.93-55.08 & 0.4393 & ~\cite{rykoff16}\\
PSZ2 G057.25-45.34 & 0.4265 & ~\cite{rykoff16}\\
PSZ2 G058.29+18.55 & 0.065 & ~\cite{PlanckSZ2}\\
PSZ2 G062.44-46.43 & 0.0909 & ~\cite{rines16}\\
PSZ2 G067.17+67.46 & 0.166 & ~\cite{rines16}\\
PSZ2 G071.63+29.78 & 0.1579 & ~\cite{rines16}\\
PSZ2 G072.62+41.46 & 0.2257 & ~\cite{rines16}\\
PSZ2 G072.79-18.73 & 0.143 & ~\cite{PlanckSZ2}\\
PSZ2 G073.97-27.82 & 0.2302 & ~\cite{rykoff16}\\
PSZ2 G080.41-33.24 & 0.1102 & ~\cite{rines16}\\
PSZ2 G081.00-50.93 & 0.2998 & ~\cite{PlanckSZ2}\\
PSZ2 G083.29-31.03 & 0.412 & ~\cite{PlanckSZ2}\\
PSZ2 G085.98+26.69 & 0.179 & ~\cite{PlanckSZ2}\\
PSZ2 G086.47+15.31 & 0.26 & ~\cite{PlanckSZ2}\\
PSZ2 G092.71+73.46 & 0.2312 & ~\cite{rines16}\\
PSZ2 G093.92+34.92 & 0.0801 & ~\cite{rines16}\\
PSZ2 G096.88+24.18 & 0.37 & ~\cite{planck_canary16}\\
PSZ2 G097.72+38.12 & 0.1709 & ~\cite{PlanckSZ2}\\
PSZ2 G098.97+24.86 & 0.0928 & ~\cite{PlanckSZ2}\\
PSZ2 G106.87-83.23 & 0.2924 & ~\cite{PlanckSZ2}\\
PSZ2 G107.10+65.32 & 0.2799 & ~\cite{PlanckSZ2}\\
PSZ2 G113.81+44.35 & 0.225 & ~\cite{PlanckSZ2}\\
PSZ2 G124.20-36.48 & 0.1971 & ~\cite{PlanckSZ2}\\
PSZ2 G125.71+53.86 & 0.3019 & ~\cite{PlanckSZ2}\\
PSZ2 G139.18+56.37 & 0.322 & ~\cite{PlanckSZ2}\\
PSZ2 G149.75+34.68 & 0.1818 & ~\cite{PlanckSZ2}\\
PSZ2 G157.43+30.34 & 0.407 & ~\cite{barrena18}\\
PSZ2 G159.91-73.50 & 0.206 & ~\cite{PlanckSZ2}\\
PSZ2 G164.18-38.88 & 0.0739 & ~\cite{PlanckSZ2}\\
PSZ2 G166.09+43.38 & 0.2173 & ~\cite{rines16}\\
PSZ2 G167.67+17.63 & 0.174 & ~\cite{PlanckSZ2}\\
PSZ2 G171.98-40.66 & 0.272 & ~\cite{planck_canary16}\\
PSZ2 G180.25+21.03 & 0.546 & ~\cite{PlanckSZ2}\\
PSZ2 G182.42-28.28 & 0.0882 & ~\cite{PlanckSZ2}\\
PSZ2 G182.59+55.83 & 0.2041 & ~\cite{rines16}\\
PSZ2 G186.37+37.26 & 0.2812 & ~\cite{rines16}\\
PSZ2 G195.60+44.06 & 0.293 & ~\cite{mehrtens16}\\
PSZ2 G195.75-24.32 & 0.203 & ~\cite{PlanckSZ2}\\
PSZ2 G218.81+35.51 & 0.1751 & ~\cite{PlanckSZ2}\\
PSZ2 G225.93-19.99 & 0.46 & ~\cite{PlanckSZ2}\\
PSZ2 G226.16-21.95 & 0.0989 & ~\cite{PlanckSZ2}\\
PSZ2 G226.18+76.79 & 0.1412 & ~\cite{rines16}\\
PSZ2 G228.16+75.20 & 0.545 & ~\cite{PlanckSZ2}\\
PSZ2 G227.61+54.87 & 0.3188 & ~\cite{PlanckSZ2}\\
PSZ2 G229.23-17.23 & 0.171 & ~\cite{PlanckSZ2}\\
PSZ2 G229.93+15.30 & 0.0704 & ~\cite{PlanckSZ2}\\
PSZ2 G236.92-26.65 & 0.1483 & ~\cite{PlanckSZ2}\\
PSZ2 G241.76-30.88 & 0.2708 & ~\cite{PlanckSZ2}\\
PSZ2 G241.79-24.01 & 0.1392 & ~\cite{PlanckSZ2}\\
PSZ2 G241.98+14.87 & 0.1687 & ~\cite{PlanckSZ2}\\
PSZ2 G244.37-32.15 & 0.2839 & ~\cite{PlanckSZ2}\\
PSZ2 G244.71+32.50 & 0.1535 & ~\cite{PlanckSZ2}\\
PSZ2 G247.19-23.31 & 0.152 & ~\cite{PlanckSZ2}\\
PSZ2 G249.91-39.86 & 0.1501 & ~\cite{PlanckSZ2}\\
PSZ2 G250.89-36.24 & 0.2 & ~\cite{PlanckSZ2}\\
PSZ2 G252.99-56.09 & 0.0752 & ~\cite{PlanckSZ2}\\
PSZ2 G253.48-33.72 & 0.191 & ~\cite{PlanckSZ2}\\
PSZ2 G256.53-65.70 & 0.2195 & ~\cite{PlanckSZ2}\\
PSZ2 G257.32-22.19 & 0.203 & ~\cite{PlanckSZ2}\\
PSZ2 G259.98-63.43 & 0.2836 & ~\cite{PlanckSZ2}\\
PSZ2 G262.27-35.38 & 0.2952 & ~\cite{PlanckSZ2}\\
PSZ2 G262.73-40.92 & 0.4224 & ~\cite{bayliss16}\\
PSZ2 G263.14-23.41 & 0.226 & ~\cite{bleem15}\\
PSZ2 G263.68-22.55 & 0.1644 & ~\cite{PlanckSZ2}\\
PSZ2 G266.04-21.25 & 0.2965 & ~\cite{PlanckSZ2}\\
PSZ2 G269.31-49.87 & 0.0853 & ~\cite{PlanckSZ2}\\
PSZ2 G271.18-30.95 & 0.376 & ~\cite{bleem15}\\
PSZ2 G271.53-56.57 & 0.3 & ~\cite{PlanckSZ2}\\
PSZ2 G272.08-40.16 & 0.0589 & ~\cite{PlanckSZ2}\\
PSZ2 G277.76-51.74 & 0.438 & ~\cite{PlanckSZ2}\\
PSZ2 G278.58+39.16 & 0.3075 & ~\cite{PlanckSZ2}\\
PSZ2 G280.17+47.83 & 0.1557 & ~\cite{PlanckSZ2}\\
PSZ2 G286.39+64.06 & 0.2287 & ~\cite{rykoff16}\\
PSZ2 G283.16-22.91 & 0.45 & ~\cite{PlanckSZ2}\\
PSZ2 G284.41+52.45 & 0.4414 & ~\cite{PlanckSZ2}\\
PSZ2 G284.97-23.69 & 0.39 & ~\cite{PlanckSZ2}\\
PSZ2 G285.63-17.23 & 0.35 & ~\cite{PlanckSZ2}\\
PSZ2 G286.62-31.24 & 0.21 & ~\cite{PlanckSZ2}\\
PSZ2 G286.98+32.90 & 0.39 & ~\cite{PlanckSZ2}\\
PSZ2 G288.63-37.66 & 0.127 & ~\cite{PlanckSZ2}\\
PSZ2 G292.56+21.97 & 0.3 & ~\cite{PlanckSZ2}\\
PSZ2 G294.68-37.01 & 0.2742 & ~\cite{PlanckSZ2}\\
PSZ2 G304.65-31.66 & 0.1934 & ~\cite{PlanckSZ2}\\
PSZ2 G304.84-41.40 & 0.41 & ~\cite{PlanckSZ2}\\
PSZ2 G306.66+61.06 & 0.0843 & ~\cite{rines16}\\
PSZ2 G306.77+58.61 & 0.0845 & ~\cite{PlanckSZ2}\\
PSZ2 G308.33-20.21 & 0.48 & ~\cite{PlanckSZ2}\\
PSZ2 G313.33+61.13 & 0.1842 & ~\cite{rines16}\\
PSZ2 G313.88-17.11 & 0.153 & ~\cite{PlanckSZ2}\\
PSZ2 G318.14-29.57 & 0.217 & ~\cite{PlanckSZ2}\\
PSZ2 G321.98-47.96 & 0.094 & ~\cite{PlanckSZ2}\\
PSZ2 G324.54-44.97 & 0.0951 & ~\cite{PlanckSZ2}\\
PSZ2 G332.23-46.37 & 0.098 & ~\cite{PlanckSZ2}\\
PSZ2 G332.87-19.26 & 0.147 & ~\cite{PlanckSZ2}\\
PSZ2 G335.58-46.44 & 0.076 & ~\cite{PlanckSZ2}\\
PSZ2 G336.60-55.43 & 0.0965 & ~\cite{PlanckSZ2}\\
PSZ2 G337.14-25.98 & 0.26 & ~\cite{PlanckSZ2}\\
PSZ2 G342.33-34.93 & 0.232 & ~\cite{PlanckSZ2}\\
PSZ2 G347.17-27.36 & 0.2371 & ~\cite{PlanckSZ2}\\
PSZ2 G349.46-59.95 & 0.351 & ~\cite{bleem15}\\
\hline 
\end{longtable}

\newpage

\begin{figure*}
    %\centering
   \includegraphics[width=17cm, height=10cm]{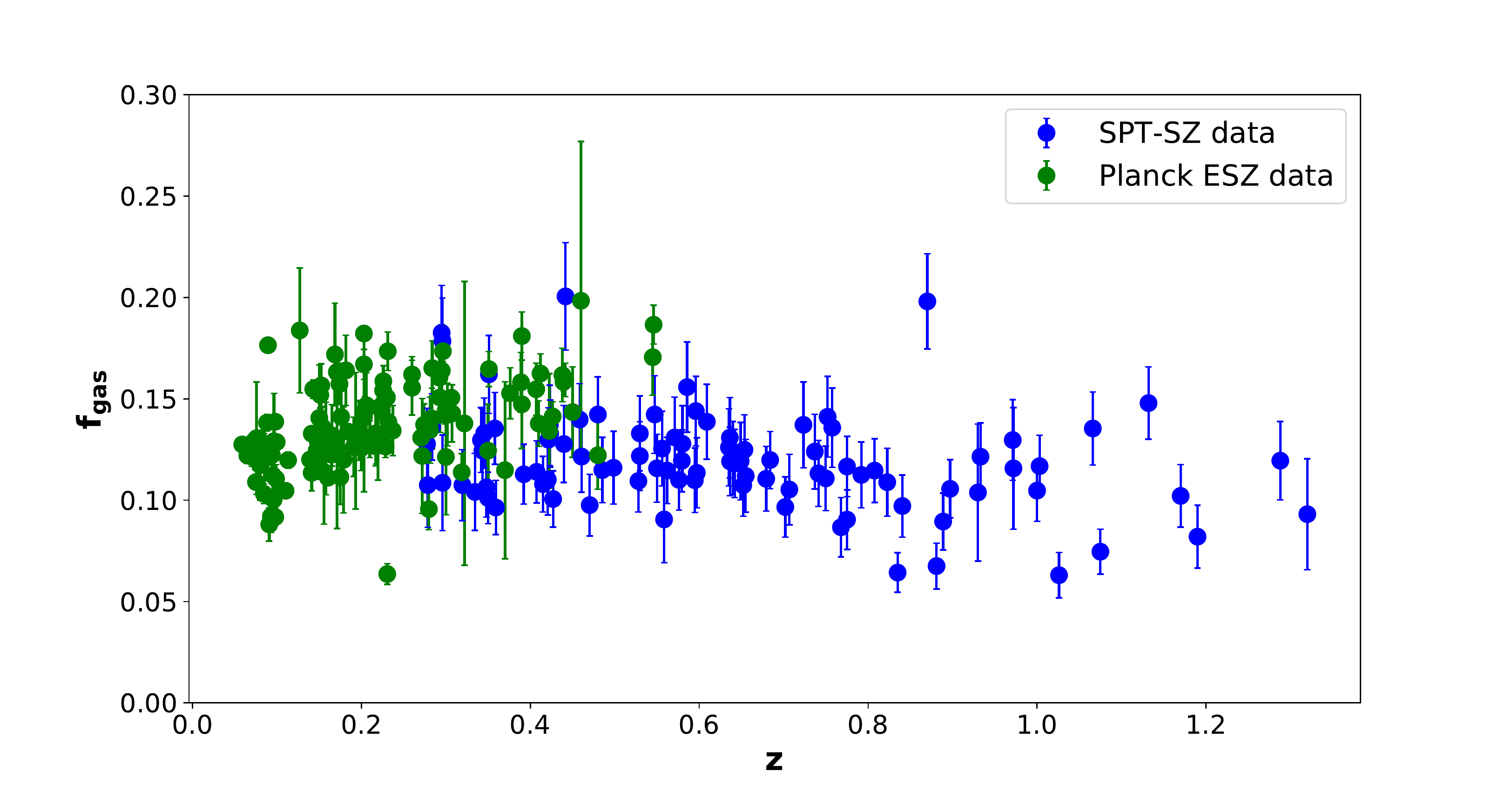} 
    \caption{$f_{gas}$ data of 94 SPT-SZ clusters sample taken from~\cite{chiu18,chiu16} and 120 Planck ESZ galaxy clusters sample taken from~\cite{lovisari20}  as a function of $z$.}
    \label{fig:f1}
\end{figure*}

The gas density was obtained from the X-ray surface brightness profile in the 0.3-2 keV energy band after fitting a double-$\beta$ model~\cite{lovisari17}. The temperatures were obtained by fitting the X-ray spectra to a thermal plasma emission model~\cite{lovisari17}. The gas mass $M_{gas}$ was then  estimated by integrating the density profile upto  $R_{500}$ after assuming spherical symmetry.
%\begin{equation}
%    M_{gas} =    \int_{0}^{R_{500}}{4\pi r^2 %\rho_{gas}(r) dr}
%    \label{eq:mgas}
%\end{equation}
Both $M_{gas}$ and $M_{500}$, along with their errors are tabulated in  ~\citet{lovisari20}.
Thereafter, their ratio $M_{gas}/M_{500}$ gives the desired gas mass fraction $f_{gas}$ values for each cluster. Fig.~\ref{fig:f1} shows the derived $f_{gas}$ values for Planck ESZ clusters as a function of redshift.

%\begin{figure}[t]
    %\centering
%    \includegraphics[width=10cm, height=8cm]{Figure_1.pdf} 
%    \caption{$f_{gas}$ data of 94 SPT-SZ clusters sample as a function of $z$.}
%    \label{fig:f1}
%\end{figure}

%\begin{figure}[t]
    %\centering
 %  \includegraphics[width=10cm, height=8cm]{Figure_2.pdf} 
 %   \caption{$f_{gas}$ data of 120 Planck ESZ galaxy clusters sample taken from~\cite{chiu18}.   }
 %   \label{fig:f2}
%\end{figure}

\section{Method}
\label{sec:method}
The gas mass fraction in galaxy clusters has been used as a cosmological probe. The expression for  $f_{gas}$ with respect  to a reference cosmological model is given by~\cite{allen08}:

\begin{equation}
    f_{gas}^{ref}= K(z)A(z)\gamma(z)\left(\frac{\Omega_b}{\Omega_m}\right) {\left[\frac{D_A^{ref}(z)}{D_A(z)}\right]}^{3/2}
	\label{eq:fgas}
\end{equation}
where $\Omega_b$ and $\Omega_m$ denote the cosmic baryon and matter fraction density (obtained from Planck 18 Cosmological analysis~\cite{planck18}) scaled by the critical density $\rho_c$; $K(z)$ = $1.0\pm0.1$~\cite{allen08} is the instrument calibration constant, which also accounts for any bias in mass due to non-thermal pressure and bulk motions in baryonic gas; $A(z)$ represents the angular correction factor which is almost in all cases, close to unity, and hence can be neglected.  The terms in the square parenthesis denote the variation in $f_{gas}$ value as the underlying cosmology is changed. $D_{A}$ is the angular diameter distance and $ref$ indicates the fiducial cosmology. For our  reference cosmology, we have used a  flat $\Lambda$CDM cosmology ($\Omega_m$ = 0.3 and $H_0$ = 70 km/sec/Mpc~\cite{lovisari20}).

Assuming flat $\Lambda$CDM cosmology, the angular diameter distance for a reference cosmology can be calculated as follows~\cite{Hogg}:
\begin{equation}
    D_{A}^{ref} = \frac{c}{H_0} \left(\frac{1}{1+z}\right) \int_{0}^{z}\frac{dz^{'}}{\sqrt{\Omega_m{(1+z^{'})^3}+(1 - \Omega_m)}}
   \label{eq:ADD}
\end{equation}
From Eq.~\ref{eq:fgas}, we can write the gas depletion factor, $\gamma(z)$ as follows: 
\begin{equation}
\label{eq:gamma_obs}
    \gamma(z)= \frac{f_{gas}^{ref}}{ A(z) K(z)} \left(\frac{\Omega_m}{\Omega_b}\right) {\left[\frac{D_A(z)}{D_A^{ref}(z)}\right]}^{3/2}
\end{equation}
This equation shows that $\gamma(z)$ is sensitive to  measurements of angular diameter distance.
In this work, similar to Z19 we have used Gaussian Process Regression to derive the angular diameter distances  by using $H(z)$ data from cosmic chronometers, which are agnostic to the underlying cosmology. $D_A(z)$ is given by
\begin{equation}
D_{A} (z) =  \left(\frac{c}{1+z}\right)\int_{0}^{z}\frac{dz^{'}}{H(z')}
\label{eq:daz}
\end{equation}
where $H(z')$ is the non-parametric  estimate of the Hubble parameter, whose determination will be described in the next section.

\begin{figure*}[t]
    %\centering
    \includegraphics[width=17cm, height=10cm]{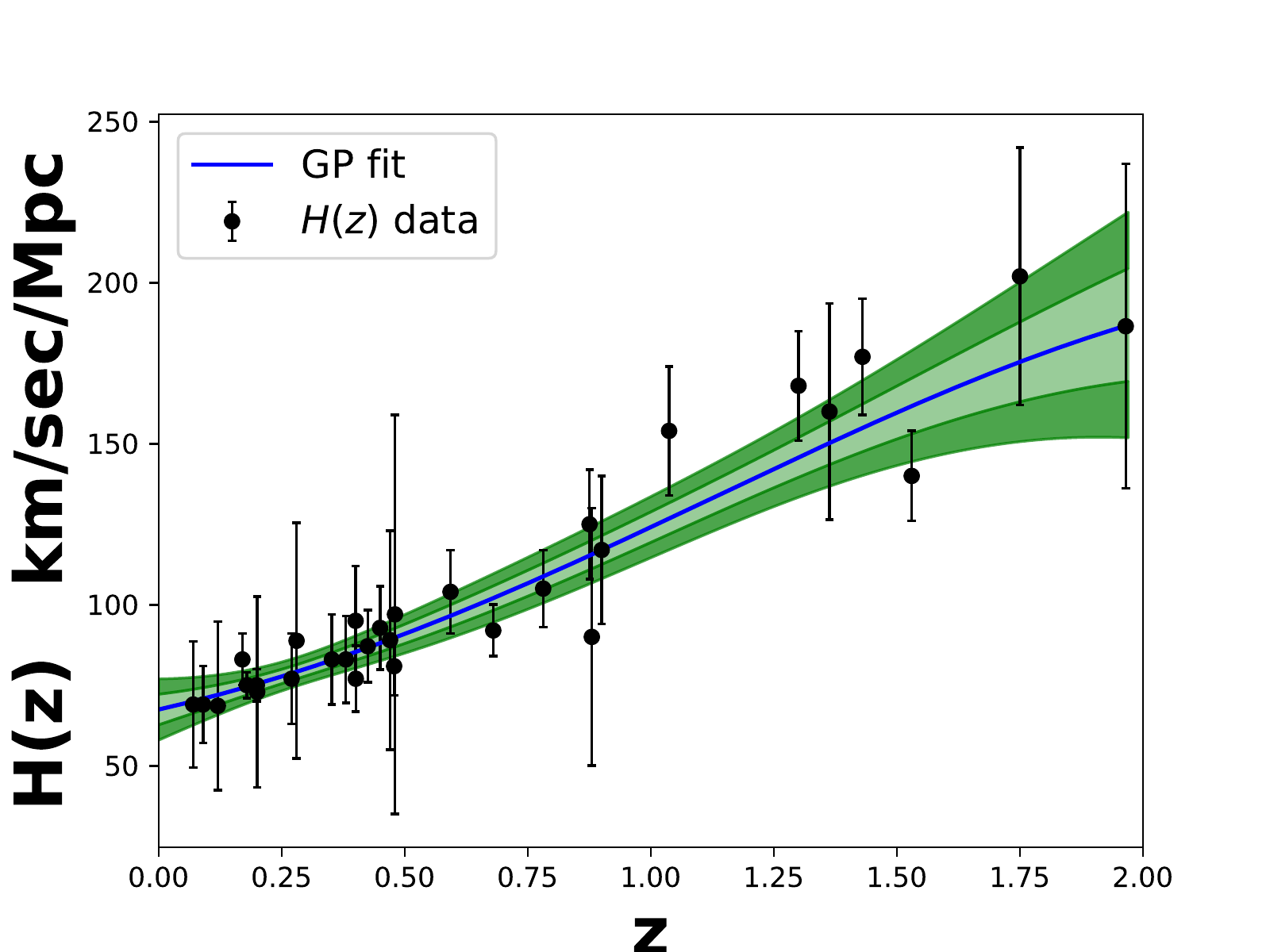}
    \caption{Reconstruction of $H(z)$ using Gaussian Processes Regression. The data points represent the 31 $H(z)$ cosmic chronometer measurements taken from ~\citet{li19}. The blue line indicates the best GP fit to data along with $1\sigma$ and $2\sigma$ error bands shown by two different shades of green bands.}
    \label{fig:f3}
\end{figure*}

\section{Analysis and results}
\label{sec:analysis}
Cosmic chronometers provide a model-independent
measurement of $H(z)$, based on the age difference between two passively evolving galaxies~\cite{Jimenez}. Therefore, they have been widely used for a variety of tests of the  standard cosmological model, as well as a whole suite of  cosmological measurements~\cite{Seikel,Meliachrono,ChenRatra,Amendola,Yang19,Haridasu2,Jesus,Rana,Ratra13,Moresco_2016,HS}. For this work, we used cosmic chronometer data to provide a non-parametric estimate of the expansion history, needed to evaluate Eq.~\ref{eq:daz}. 

Gaussian Processes (GPs) extend the idea of Gaussian distribution, and are characterized by the mean function and the covariance (kernel) function. They offer a non-parametric way to model a function~\cite{Seikel}. In this work, we choose a squared exponential (RBF) kernel function which is defined as,
\begin{equation}
 \label{eq:kernal}
  K(x,\Tilde{x}) = \sigma_f^2 exp  {\left[\frac{-(x-\Tilde{x})^2}{2l^2}\right]}, 
\end{equation}
where $\sigma_f^2$ and $l$ are the hyperparameters of the kernel function. The length parameter $l$ controls the smoothness of the function. To implement the GPs, we used the GaPP (Gaussian Process in Python) code~\cite{GaPP}. 
%SD: please write a few sentences about Gaussian processes including which kernel and code were used to calculate GPR

Following Z19, we used the 31 $H(z)$ cosmic chronometer measurements from~\citet{li19} (same  as that  used in ~\cite{HS}) in the redshift range $0.07 \leqslant z \leqslant 1.965$
to obtain non-parametric estimates of $H(z)$ at any redshift. 
Fig.~\ref{fig:f3} shows the GP reconstructed $H(z)$ from cosmic chronometers along with  $1\sigma$ and $2\sigma$ uncertainties.
We then reconstruct the GP reconstructed $H(z)$ values for SPT-SZ and  Planck ESZ  redshifts in order to obtain the angular diameter distances. Thereafter, these $H(z)$ values were used to derive the angular diameter distances $D_A$ using 
Eq.~\ref{eq:daz}. $D_A^{ref}$ is calculated from Eq.~\ref{eq:ADD} by assuming  the cosmological parameters outlined in Sect.~\ref{sec:method}.

\begin{figure}[t]
    \centering
    \includegraphics[width=10cm,height=8cm]{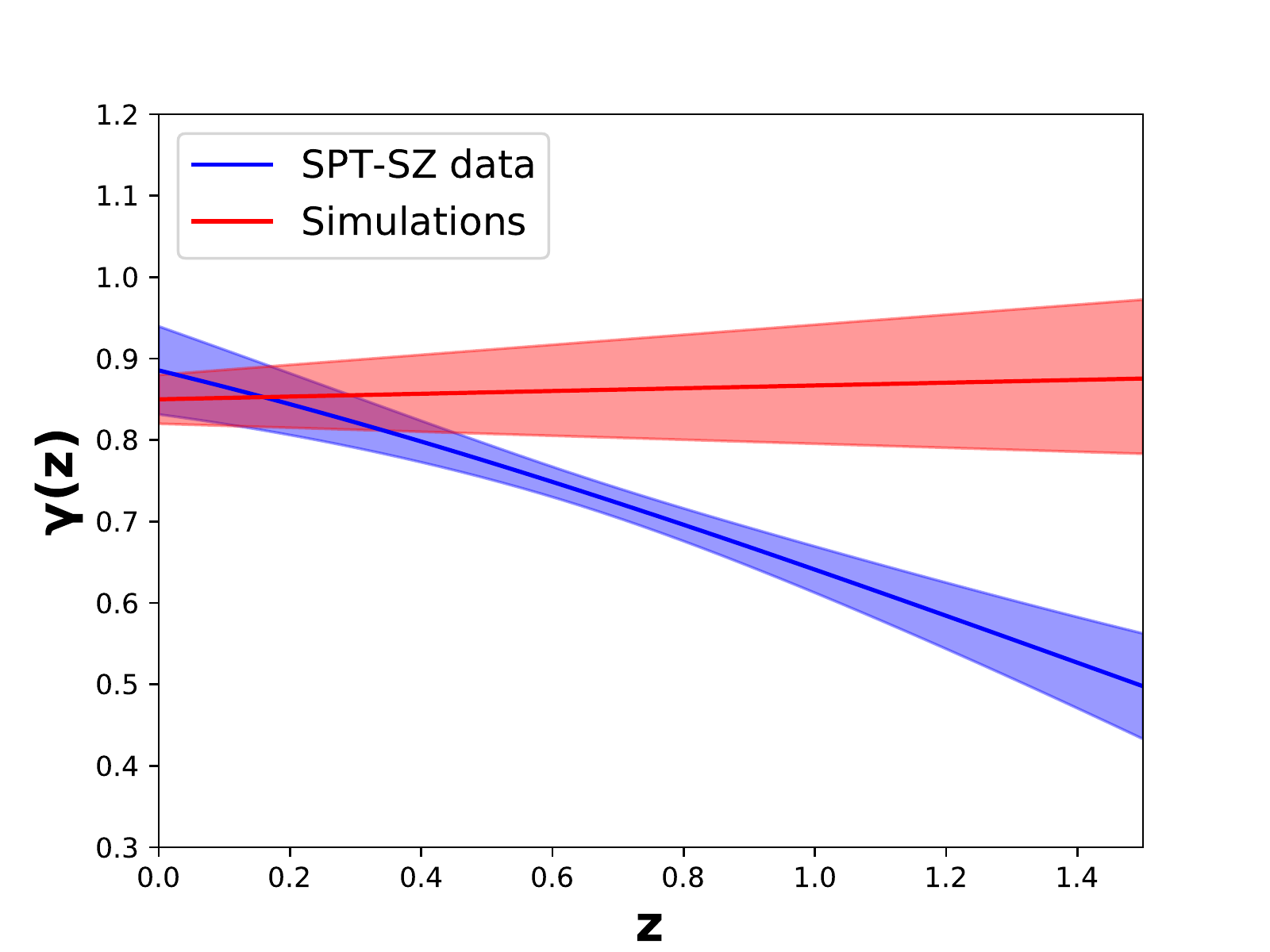} 
    \caption{Reconstructed $\gamma(z)$ as a function of $z$. The blue line represents the best GP reconstructed fit for the SPT-SZ data along with the $1\sigma$ error band (shown by blue shaded region). The red band shows the NR simulation results at $R_{500}$ from~\cite{planelles13} by assuming the Eq.~\ref{eq:gammavariation}. }
    \label{fig:f4}
\end{figure}

\begin{figure}[t]
    \centering
    \includegraphics[width=10cm, height=8cm]{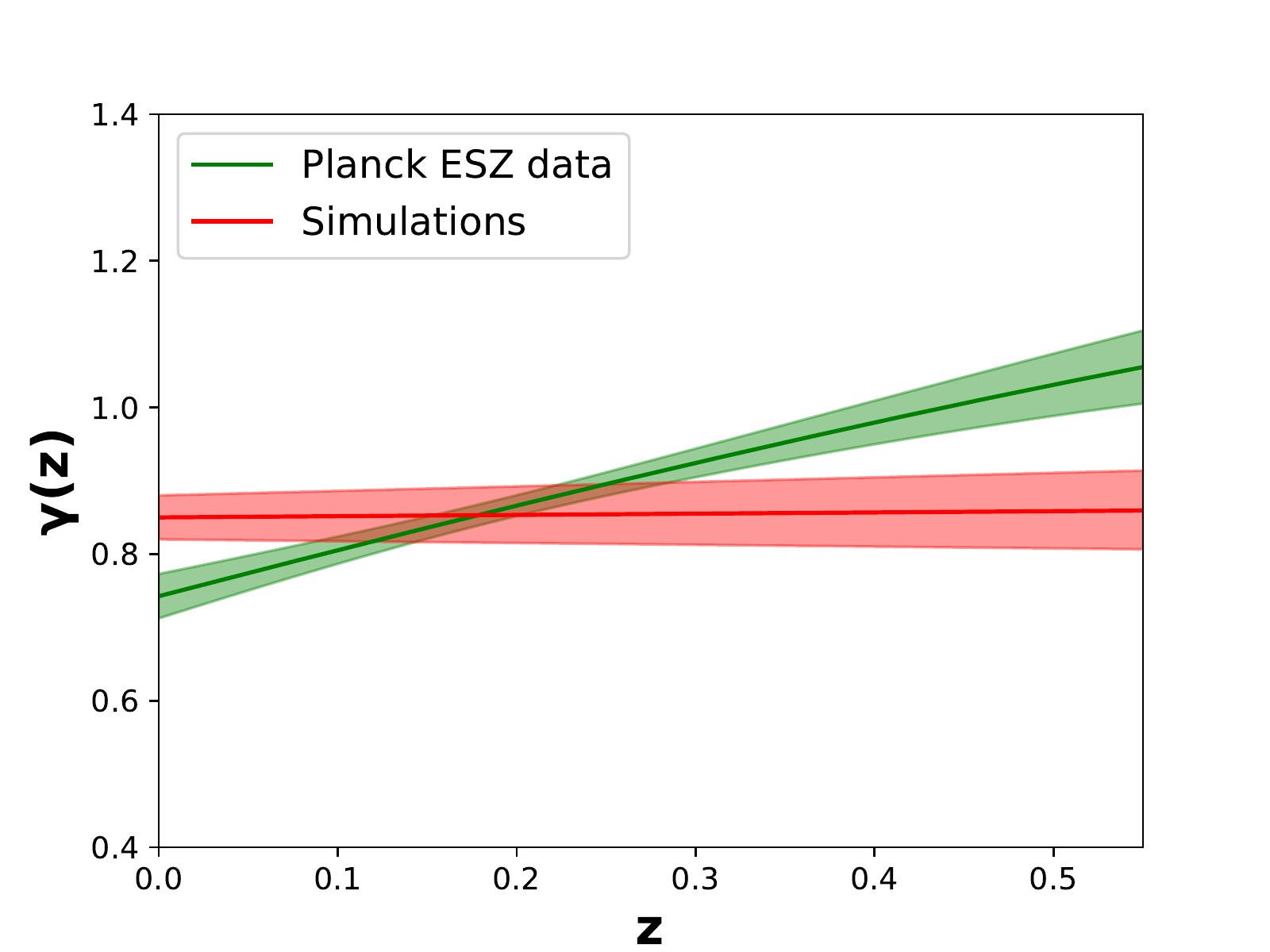} 
    \caption{Reconstructed $\gamma(z)$ as a function of $z$. The green line represents the best GP reconstructed fit for the Planck ESZ data along with the $1\sigma$ error band(shown by green shaded region). The red region shows the NR simulation results at $R_{500}$ from~\cite{planelles13} by assuming the Eq.~\ref{eq:gammavariation}.}
    \label{fig:f5}
\end{figure}

Finally, we reconstructed the gas depletion factor $\gamma (z)$  using Gaussian Process, after incorporating all  the other factors presented in  Eq.~\ref{eq:gamma_obs}. We also propagated the errors in $\Omega_b$, $\Omega_M$, $K$, angular diameter distance,  and redshifts, if they were provided.
Figs.~\ref{fig:f4} and ~\ref{fig:f5} display our reconstructed $\gamma(z)$ as a function of $z$ for both SPT-SZ and Planck ESZ data, respectively. Both these data  show opposite  trends for the  gas depletion factor with redshift. We find a decreasing  $\gamma$ for  SPT-SZ data, whereas for Planck ESZ data, a slightly increasing trend is observed. For comparison,~\citet{planelles13} reported a negligible evolution of gas depletion factor by using Eq.~\ref{eq:gammavariation} for their NR simulations. They obtained $\gamma_0 = 0.85\pm0.03$ and  $\gamma_1 = 0.02\pm 0.05$ at $R_{500}$. The simulation results can be found in the red shaded region in  Figs.~\ref{fig:f4} and  ~\ref{fig:f5}. \rthis{Therefore, at face value, our results do not agree with these hydrodynamical simulations based on $\Lambda$CDM.}

To confirm and quantify the significance of the trends seen in  $\gamma(z)$, we also did a  parametric  fit of our estimated depletion factor measurements  to Eq.~\ref{eq:gammavariation}.
We maximize the following likelihood function $\mathcal{L}$ in our analysis, where
%\begin{widetext}
\begin{equation}
    \label{eq:logL}
   -2\ln\mathcal{L} = \sum_{obs} \ln 2\pi{\sigma_{obs}^2}+ \sum_{i=1}^{n}\frac{(\gamma_{model}(z_i)-\gamma_{obs}(z_i))^2}{\sigma_{i,obs}^2} , 
\end{equation}          %\end{widetext}  
where $\gamma_{model}$ is defined in Eq.~\ref{eq:gammavariation}, $\gamma_{obs}$ is obtained from Eq.~\ref{eq:gamma_obs}, and  $\sigma_{i,obs}^2$ represents the error  in $\gamma_{obs}$.  $\sigma_{i,obs}$ is calculated by propagating the errors in $f_{gas}^{ref}$, $K$, $\Omega_m$,  cluster $z$, $\Omega_b$, $D_A(z)$, and $D_A^{ref}$. In addition to these observational errors, $\sigma_{obs}$ also includes an unknown  intrinsic scatter ($\sigma_{int}$) added in quadrature, similar to our recent works on galaxy cluster scaling relations~\cite{Gopika,Bora,Pradyumna}.

\begin{figure*}[h]
    %\centering
 \includegraphics[width=10cm, height=10cm]{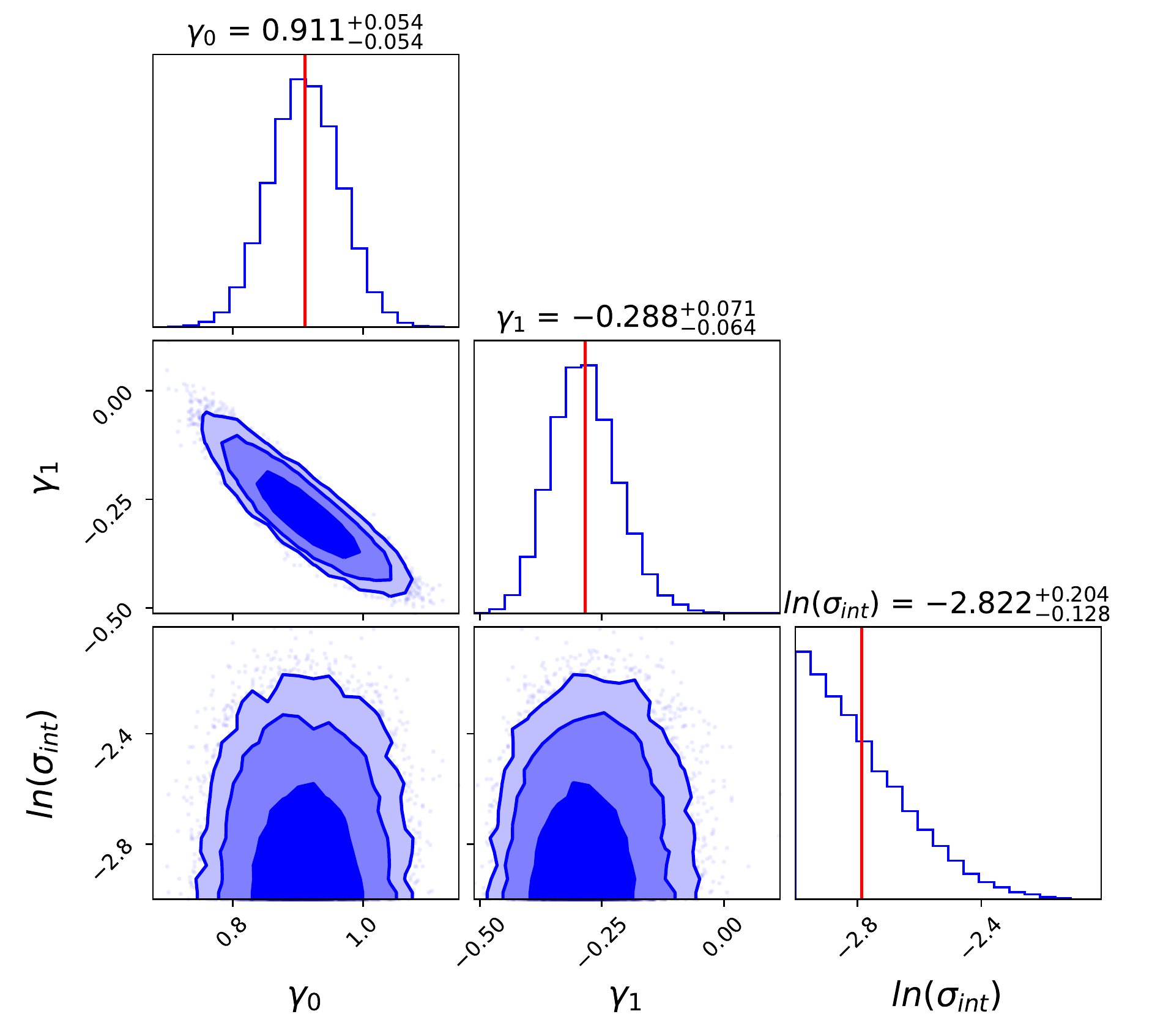}    \caption{\textbf{For SPT-SZ dataset:} Constraints on the parameters $\gamma_0$ and $\gamma_1$ along with $\ln(\sigma_{int})$, defined in Eq.~\ref{eq:logL}. The plots along the diagonal are the one-dimensional marginalized likelihood distributions. The contour plots represents the two-dimensional marginalized constraints showing the 68\%, 95\%, and 99\% credible regions. These contours have been obtained using the {\tt Corner} python module~\cite{corner}. }
 \label{fig:f6}
\end{figure*}

\begin{figure*}[h]
    %\centering
 \includegraphics[width=10cm, height=10cm]{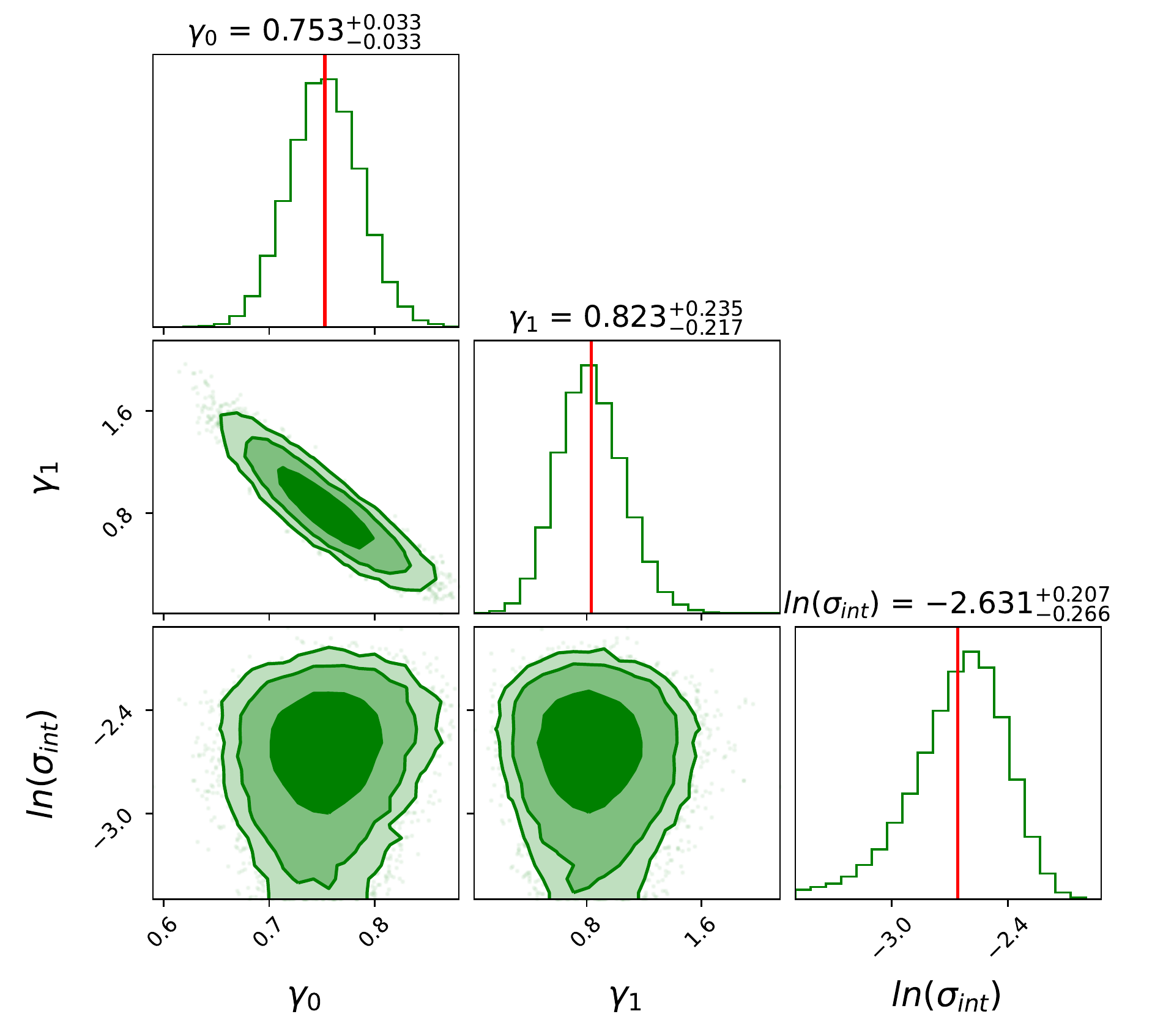}    \caption{\textbf{For Planck ESZ dataset:} Constraints on the parameters $\gamma_0$ and $\gamma_1$ along with $\ln(\sigma_{int})$. The plots along the diagonal are the one-dimensional marginalized likelihood distributions. The contour plot represents the two-dimensional marginalized constraints showing the 68\%, 95\%, and 99\% credible regions. These contours have been obtained using the {\tt Corner} python module~\cite{corner}. }
    \label{fig:f7}
\end{figure*}

To estimate the model parameters($\gamma_0$ and $\gamma_1$), we maximize the likelihood using the {\tt emcee} MCMC sampler~\cite{emcee}.
For both the datasets, we adopt uniform priors on $\gamma_0$  and $\gamma_1$: $-1.5 \leqslant \gamma_0 \leqslant 1.5$ and $-2.0 \leqslant \gamma_1 \leqslant 3.0$. For the intrinsic scatter, similar to ~\cite{Tian}, we assume log-uniform priors between $10^{-5}$ and 0.1. \rthis{The log-uniform prior ensures that the intrinsic scatter is greater than zero.}
The 68\%, 95\%, and 99\% 2-D marginalized credible intervals, along with the marginalized one-dimensional likelihoods for each parameters  are displayed in Fig.~\ref{fig:f6} and Fig.~\ref{fig:f7} respectively. 

From this analysis, we get $\gamma_0 = 0.915\pm0.050$ and $\gamma_1 = -0.305_{-0.058}^{+0.064}$ for SPT-SZ data, and  $\gamma_0 = 0.741\pm0.029$ and $\gamma_1 = 0.834_{-0.199}^{+0.214}$ for Planck ESZ data. The SPT-SZ data results indicate a decreasing  trend of $\gamma$(z) with redshift at about 5$\sigma$ significance. On the other hand, the Planck ESZ results show an increasing $\gamma$ as a function of redshift at about $4\sigma$. Both the results contradict those  from hydrodynamical simulations ~\cite{planelles13}. Table~\ref{tab:summary_table} summarizes the constraints on parameters $\gamma_0$ and $\gamma_1$,  which we have obtained from the datasets, along with a summary of previous compilations in literature using both data and simulations. We note that among the  previous observational results, only Z19 had found $\gamma(z)$ decreasing with redshift at about $1.6\sigma$.

%See the appendix to check out the reconstruction of Fig [~\ref{fig:f4},~\ref{fig:f5},~\ref{fig:f6},~\ref{fig:f7}] by using  {\tt scikit-learn} ~\cite{scikit-learn}.

\begin{table*}
%	\centering
	\caption{Constraints on parameters $\gamma_0$ and $\gamma_1$ (cf. Eq.~\ref{eq:gammavariation}) from different estimates in literature (including this work). The last two rows show the gas depletion factors for a subset of SPT-SZ  and Planck ESZ clusters within the same redshift range. }
	\label{tab:summary_table}
	\begin{tabular}{|l|c|c|c|c|r|} % six columns, alignment for each
		\hline
		\boldmath{$f_{gas}$} & \textbf{Redshift range} & \textbf{Cluster Radius} &  \boldmath$\gamma_0$ & \boldmath$\gamma_1$ & \textbf{Reference}\\
		\hline
	
	    NR Simulation & $0.0  \leqslant z  \leqslant 1.0$ & $R_{200}$ & $0.86\pm0.02$ & $0.00\pm0.04$ & ~\cite{planelles13}  \\
	    NR Simulation & $0.0  \leqslant z  \leqslant 1.0$ & $R_{500}$ & $0.85\pm0.03$ & $0.02\pm0.05$ &  ~\cite{planelles13} \\
	    NR Simulation & $0.0  \leqslant z  \leqslant 1.0$ & $R_{2500}$ & $0.79\pm0.07$ & $0.07\pm0.12$ & ~\cite{planelles13} \\
	    CSF Simulation & $0.0  \leqslant z  \leqslant 1.0$ & $R_{200}$ & $0.68\pm0.03$ & $0.00\pm0.06$ & ~\cite{planelles13}  \\
	    CSF Simulation & $0.0  \leqslant z  \leqslant 1.0$ & $R_{500}$ & $0.63\pm0.03$ & $0.01\pm0.08$ &  ~\cite{planelles13} \\
	    CSF Simulation & $0.0  \leqslant z  \leqslant 1.0$ & $R_{2500}$ & $0.49\pm0.06$ & $-0.04\pm0.18$ & ~\cite{planelles13} \\
	    AGN Simulation & $0.0  \leqslant z  \leqslant 1.0$ & $R_{200}$ & $0.75\pm0.03$ & $-0.03\pm0.05$ & ~\cite{planelles13}  \\
	    AGN Simulation & $0.0  \leqslant z  \leqslant 1.0$ & $R_{500}$ & $0.71\pm0.03$ & $-0.03\pm0.06$ &  ~\cite{planelles13} \\
	    AGN Simulation & $0.0  \leqslant z  \leqslant 1.0$ & $R_{2500}$ & $0.55\pm0.07$ & $-0.04\pm0.18$ & ~\cite{planelles13} \\
	    \hline
	    $f_{gas}$/SNIa & $ 0.078  \leqslant z  \leqslant 1.063$ & $R_{2500}$ & $0.86\pm0.04$ & $-0.04\pm0.12$ & ~\cite{holanda1706} \\
	    \hline
	    $f_{gas}$/Cluster I & $0.14  \leqslant z  \leqslant 0.89$ & $R_{2500}$ & $0.76\pm0.14$ & $-0.42_{-0.40}^{+0.42}$ & ~\cite{holanda1711} \\
	     $f_{gas}$/Cluster II & $0.14  \leqslant z  \leqslant 0.89$ & $R_{2500}$ & $0.72\pm0.01$ & $0.16\pm0.36$ &  ~\cite{holanda1711} \\
	     $f_{gas}$/$\Lambda$CDM & $0.12  \leqslant z  \leqslant 0.78$ & $R_{2500}$ & $0.84\pm0.07$ & $-0.02\pm0.14$ & ~\cite{holanda1711} \\
	     \hline
		ACTPol & $0.1  \leqslant z  \leqslant 1.4$ & $R_{500}$ & $0.840\pm0.025$ & $-0.072_{-0.049}^{+0.044}$ & ~\cite{zheng19} \\
	    ACTPol-re & $0.1  \leqslant z  \leqslant 1.0$  & $R_{500}$ &  $0.835\pm0.028$ & $-0.060_{-0.063}^{+0.056}$ & ~\cite{zheng19} \\
	    \hline 
		\textbf{SPT-SZ} & $0.278  \leqslant z  \leqslant 1.32$  & $R_{500}$ & $\mathbf{0.911\pm0.05} $ & $\mathbf{-0.288_{-0.06}^{+0.07}} $ & \textbf{This work} \\
		\textbf{Planck ESZ} & $0.059  \leqslant z  \leqslant 0.546$  & $R_{500}$ &  $\mathbf{0.753\pm0.03}$ & $\mathbf{0.823^{+0.24}_{-0.22}} $ & \textbf{This work} \\
		\hline 
		\textbf{SPT-SZ (subset)} & $0.278  \leqslant z  \leqslant 0.546$  & $R_{500}$ & $\mathbf{0.537\pm0.16} $ & $\mathbf{1.02_{-0.79}^{+1.44}} $ & \textbf{This work} \\
		\textbf{Planck-ESZ (subset)} & $0.278  \leqslant z  \leqslant 0.546$  & $R_{500}$ & $\mathbf{ 0.884\pm0.17 } $ & $\mathbf{0.24_{-0.47}^{+0.7}} $ & \textbf{This work}
		\\ \hline
	\end{tabular}
\end{table*}

\section{Discussions}  
\label{sec:dicussions}
Given the contradictory results for $\gamma(z)$ between the SPT-SZ and Planck-ESZ datasets as well as with previous results in literature, we carry out various sanity checks of our results along with  extensions of tests carried  out earlier, to verify the robustness of our conclusions, and to see if there is one deciding parameter, which governs the difference between the results.

\subsection{Comparison with C18}
A test of the variation of the gas mass fraction with redshift  (using different parametric forms) for the SPT-SZ sample was also carried out in C18, and no evidence for a variation of the depletion factor for redshift was found (cf. Table 6 in C18.) At prima facie, it may seem that our results contradict those in C18. However, $\gamma(z)$ defined in this work (cf. Eq.~\ref{eq:gammavariation}) is a function of $\Omega_b$, $\Omega_M$, and $D_A(z)$, besides the gas mass fraction. On the other hand, C18 carried out a joint  test for the variation of gas mass fraction with redshift and $M_{500}$.
 The functional forms used in C18 to test for  redshift trends are $\propto (1+z)^{\alpha}$ and $(E(z))^{\gamma}$, where $E(z)$ is the cosmic expansion factor in $\Lambda$CDM~\cite{Ratra13}. Here, $\gamma(z)$ estimated in this work is proportional to $\left[D_A(z)\right]^{3/2}$ (cf. Eq.~\ref{eq:gamma_obs}), which is the integral of the reciprocal of $E(z)$.  $\gamma(z)$ studies in this work also does not have an explicit dependence on the mass. We have also reconstructed $D_A(z)$ in a non-parameteric way using cosmic chronometers.

However, in order to test the consistency of our results with C18 (for SPT-SZ), we  now check  for a  variation of only the raw $f_{gas}$ data  to a few different parametric functions (including those used in C18). 
The full list of functions used to test variations of gas mass fraction are indicated below:
\begin{eqnarray}
f_{gas} &=& f_0(1 + f_1 z)  \label{eq:9}\\
f_{gas} &=& \left(\frac{1+z}{1+z_{pivot}}\right)^{\alpha} \label{eq:10} \\
f_{gas} &=& \left(\frac{E(z)}{E(z_{pivot})}\right)^{\beta} \label{eq:11}
\end{eqnarray}
where $z_{pivot}$ is the pivot value of the redshift, equal to 0.6 for SPT and 0.2 for Planck.  With these functional forms, we remove any explicit dependence on the angular diameter distance ratio.
The best-fit values of the free parameters in the above equations for both the datasets were found by constructing a likelihood similar to Eq.~\ref{eq:logL}. These best-fit parameters can be found in Table~\ref{tab:extendedanalysis} for both SPT-SZ and Planck-ESZ. We find that for the SPT-SZ sample, the parameters $f_1$, $\alpha$, and $\beta$, which encode the dependence on redshift are consistent with no variation (within 1$\sigma$). Therefore, this agrees with the results in C18, which found no redshift dependence.

However, for the Planck sample $f_1$ in Eq.~\ref{eq:9} is equal to $0.85 \pm 0.17$, indicating a positive slope which is 5$\sigma$ discrepant from zero (which corresponds to no redshift evolution). This is one of the main determining factors for the positive $\gamma_1$ for Planck-ESZ. The deviations for $\alpha$ and $\beta$ from  a zero value are within $1-2\sigma$.

\subsection{$\gamma(z)$ within the overlapping redshift range}
The SPT sample extends upto very high redshift ($z=1.32$) compared to Planck, whose most distant cluster is located at $z=0.546$. Similarly, the Planck sample contains about 84 low redshift clusters with $z<0.278$, which is the lowest redshift for the SPT sample. We now check if the gas depletion factor for both the samples within the overlapping redshift range ($0.278\leqslant z \leqslant  0.546$)  display the same trend. We get a total of 34 SPT and 36 Planck clusters with these redshift cuts. We applied the same procedure as in Sect.~\ref{sec:analysis} to determine $\gamma_0$ and $\gamma_1$ for these sub-samples. These values can be found in the last two rows of Table~\ref{tab:summary_table}. The best-fit value of $\gamma_1$ for SPT-SZ and Planck-SZ is equal to $1.02_{-0.79}^{+1.44}$ and $0.24_{-0.47}^{+0.7}$, respectively. Therefore, we see that although there is a mild deviation of $\gamma_1$ from a zero value for SPT-SZ, their values are consistent between the two datasets within $1\sigma$.    

\subsection{Dependence on $M_{500}$}
\rthis{The fits to $f_{gas}$  done in C18 also had an explicit dependence on $M_{500}$ given by $f_{gas} \propto M_{500}^{\zeta}$.}
C18 then found $\zeta =0.33 \pm 0.07$ with a $4.5\sigma$ deviation from self-similarity (\rthis{where, self-similarity  corresponds to $\zeta=0$}) in agreement with previous works~\cite{Mohr99,Neumann01,vikhlinin06,vikhlinin09}. So far, we have not accounted for any dependence on the halo mass, while estimating the gas depletion factor as a function of redshift.

In order to check the possible dependence on the mass of the cluster,  and evaluate its impact on  the gas depletion factor, we augment Eq.~\ref{eq:gammavariation} with an additional dependence on $M_{500}$ as follows:
\begin{equation}
 \gamma(z)= \gamma_0(1+\gamma_1 z)\left(\frac{M_{500}}{M_{pivot}}\right)^{\zeta}
\label{eq:gammavariationwithmass}
\end{equation}
 where $M_{pivot}$ corresponds \rthis{to the median mass for the SPT and Planck samples.} These are equal to 
$ 4.8 \times 10^{14}M_{\odot}$  and $6.0 \times 10^{14}M_{\odot}$ for  SPT-SZ and Planck ESZ, respectively. To determine these parameters,  we use an extension of the same procedure as in Sect.~\ref{sec:analysis}, by including an additional dependence on $M_{500}$.
While constructing  the likelihood, the errors in $M_{500}$ were also included.
The marginalized posterior intervals for the gas depletion parameters along with $\zeta$, which encodes the variation with $M_{500}$, can be found in Fig.~\ref{fig:f14} and Fig.~\ref{fig:f15} for SPT-SZ and Planck-ESZ, respectively. The best-fit values for these parameters can be found in Table~\ref{tab:extendedanalysis}. For SPT-SZ, we  find that 
$\zeta= 0.257 \pm 0.09$, indicating  a 2.9$\sigma$ deviation from self-similarity. For Planck-ESZ, we have  $\zeta=0.083 \pm 0.06$, with only a 1.4$\sigma$, deviation from self-similarity.  
After allowing for this dependence on 
$M_{500}$, we find that $\gamma_1=-0.23 \pm 0.08$ and $0.57 \pm 0.26$ for SPT-SZ and Planck-ESZ respectively. This shows that the discrepancy with respect to a constant depletion factor as a function of redshift decreases to $2.9\sigma$ and $2.2\sigma$ for SPT-SZ and Planck-ESZ, respectively, as  compared to the $4-5\sigma$ significance, which we had obtained earlier without including an explicit dependence on the halo mass.
Therefore, we conclude that although the statistical significance of the rising/falling slope  as a function of redshift decreases, it does not completely go away, when we include a dependence on the total mass. Only its significance reduces to between $2-3\sigma$. \rthis{However, given that the inclusion of halo mass in our fit reduces the discrepancy with respect to no evolution of $\gamma_1$, one cannot decouple the two factors while testing for a redshift of the gas depletion factor.}

\subsection{Other possible sources of systematics}
Here, we list other possible sources of errors, which could affect our estimates of $\gamma(z)$. A detailed study of the impact of  each of these effects is beyond the scope of this work. The most crucial ingredient is the determination of cluster masses, which are obtained from the observable to halo mass scaling relations and is limited by the systematic uncertainties. The gas mass can be sensitive to the assumed temperature profile ~\cite{ettori09}. Errors due to hydrostatic equilibrium assumption could be upto 15-20\%~\cite{biffi}. Other possible sources of systematic errors in the gas mass fraction determination such as magnetic field, thermal evaporation, and non-linear translation of analysis variables are discussed in ~\cite{Loeb,Linder,Jain}. Finally, we point out that the list of SPT-SZ and Planck-SZ analyzed in this work is not the full sample. The PSZ2 catalog contains 1653  cluster candidates (with 1,203 confirmed clusters) with signal to noise ratio above 4.5~\cite{PlanckSZ2}. This work has analyzed only about 10\% of the full sample, which was followed up in X-rays at the time of Planck ESZ release. Optical and X-ray studies  of Planck SZ clusters is still in progress~\cite{Melin19,Zaznobin}. Similarly, the SPT-SZ dataset  analyzed in this work is only a quarter of the full 2500 sq. degree survey sample used for the cosmological analysis in ~\cite{bocquet19}. However, the SPT clusters are also been imaged in X-rays through dedicated Chandra~\cite{mcdonald13} and XMM~\cite{Bulbul} based follow-ups. Since our currently analyzed  subset may not be representative of the full SPT and Planck SZ sample, it would be interesting to carryout a followup analysis, once  the follow-up data for the full sample is available.

\begin{table*}[]
\caption{\label{tab:extendedanalysis}. Constraints on the  parameters  of various extended tests of  the gas depletion factors as discussed in Sect.~\ref{sec:dicussions} for  both  SPT and Planck clusters.}
    \centering
    \begin{tabular}{|l|c|c|r|} \hline
    \textbf{Parameter} & \textbf{Equation} & \textbf{SPT-SZ} &   \textbf{Planck-ESZ}\\ \hline 
      $f_0$ & Eq.~\ref{eq:9} & $0.12\pm0.006$  &  $0.11 \pm 0.004$ \\
      $f_1$ & Eq.~\ref{eq:9}     & $-0.06^{+0.084}_{-0.075}$ & $0.85^{+0.18}_{-0.17}$ \\
      $\alpha$ &  Eq.~\ref{eq:10}  & $-0.26^{+0.43}_{-0.39}$ & $-0.7^{+0.62}_{-0.61}$\\
      $\beta$ &  Eq.~\ref{eq:11}  & $-0.55^{+0.46}_{-0.48}$ & $-1.64^{+0.98}_{-1.0}$\\
      $\gamma_0$ &  Eq.~\ref{eq:gammavariationwithmass}  &  $0.86\pm 0.05$ & $0.79\pm0.04$ \\
      $\gamma_1$ &  Eq.~\ref{eq:gammavariationwithmass}  & $-0.23_{-0.07}^{+0.08}$  & $0.57_{-0.25}^{+0.27}$\\
      $\zeta$ &  Eq.~\ref{eq:gammavariationwithmass}  & $0.257\pm0.09$ & $0.083\pm0.06$ \\ \hline 
    \end{tabular}
\end{table*}
\section{Conclusions}
\label{sec:conclusion}

The gas mass fraction ($f_{gas}$)  in galaxy clusters has been used as a cosmological probe in a number of works. A key assumption in these analyses is that the gas mass fraction  (at the particular radius used for cosmological measurements)  does not vary with redshift. This has also been confirmed with  hydrodynamical simulations  of galaxy clusters which use $\Lambda$CDM as the base model,  which show no evolution of the gas depletion factor with redshift ($z< 1$)~\cite{planelles13}.  

To verify this {\it ansatz} with real data in a model-independent method without any dependence on the underlying cosmology,  Z19 carried out a model-independent test  for the variation of gas depletion factor using 182 clusters  from ACTPol. Using the parameterization in Eq.~\ref{eq:gammavariation}, they found that  $\gamma_0 = 0.840\pm0.025$ and $\gamma_1 = -0.072_{-0.049}^{+0.044}$, which indicates a decreasing trend at about $1.4\sigma$. In this work, we  apply the same procedure as Z19 using the data sets from SPT-SZ and Planck ESZ in the redshift range $0.278 \leqslant z \leqslant 1.32$ and $0.059 \leqslant z \leqslant 0.546$, respectively. One difference with respect to Z19, however is that we used direct gas mass fraction measurements 
from SPT and Planck, instead of the empirical relation between halo and gas mass used in Z19.
These measurements  were obtained using a combination of SZ data along with follow-up X-ray observations from Chandra and XMM, for SPT and Planck, respectively. 

Our  $f_{gas}$ measurements as a function of redshift can be found  in Fig.~\ref{fig:f1} for SPT-SZ and Planck ESZ clusters. We derived the angular diameter distance for both the datasets using GP fitting from $H(z)$ data reconstructed using cosmic chronometers. The reconstructed  gas depletion factor can be found in  Fig.~\ref{fig:f4} and Fig.~\ref{fig:f5} for SPT-SZ and Planck ESZ data respectively. To quantify the significance  of  our results obtained from GP reconstruction, we also did a maximum likelihood fit using Eq.~\ref{eq:gammavariation}. We find that $\gamma_0 = 0.915\pm0.05$ and $\gamma_1 = -0.305_{-0.058}^{+0.064}$ for SPT-SZ data; whereas  $\gamma_0 = 0.741\pm0.029$ and $\gamma_1 = 0.834 \pm 0.2$ for Planck ESZ data.
 The credible intervals  for  $\gamma_0$ as well as $\gamma_1$ along with marginalized 1-D distributions are shown in  Fig.~\ref{fig:f6} and Fig.~\ref{fig:f7} for SPT-SZ and Planck ESZ data respectively.   We find  $\gamma$ decreases with redshift (at about 5$\sigma$) for SPT-SZ clusters,  whereas it increases with  redshift for Planck ESZ clusters (at about 4$\sigma$). Both these findings  also contradict  hydrodynamical simulations which show a constant gas depletion factor. Previously, only Z19 had found a slightly decreasing trend  with redshift for $\gamma(z)$ using ACT-Pol data. 
 
\begin{figure*}[h]
    %\centering
 \includegraphics[width=12cm, height=10cm]{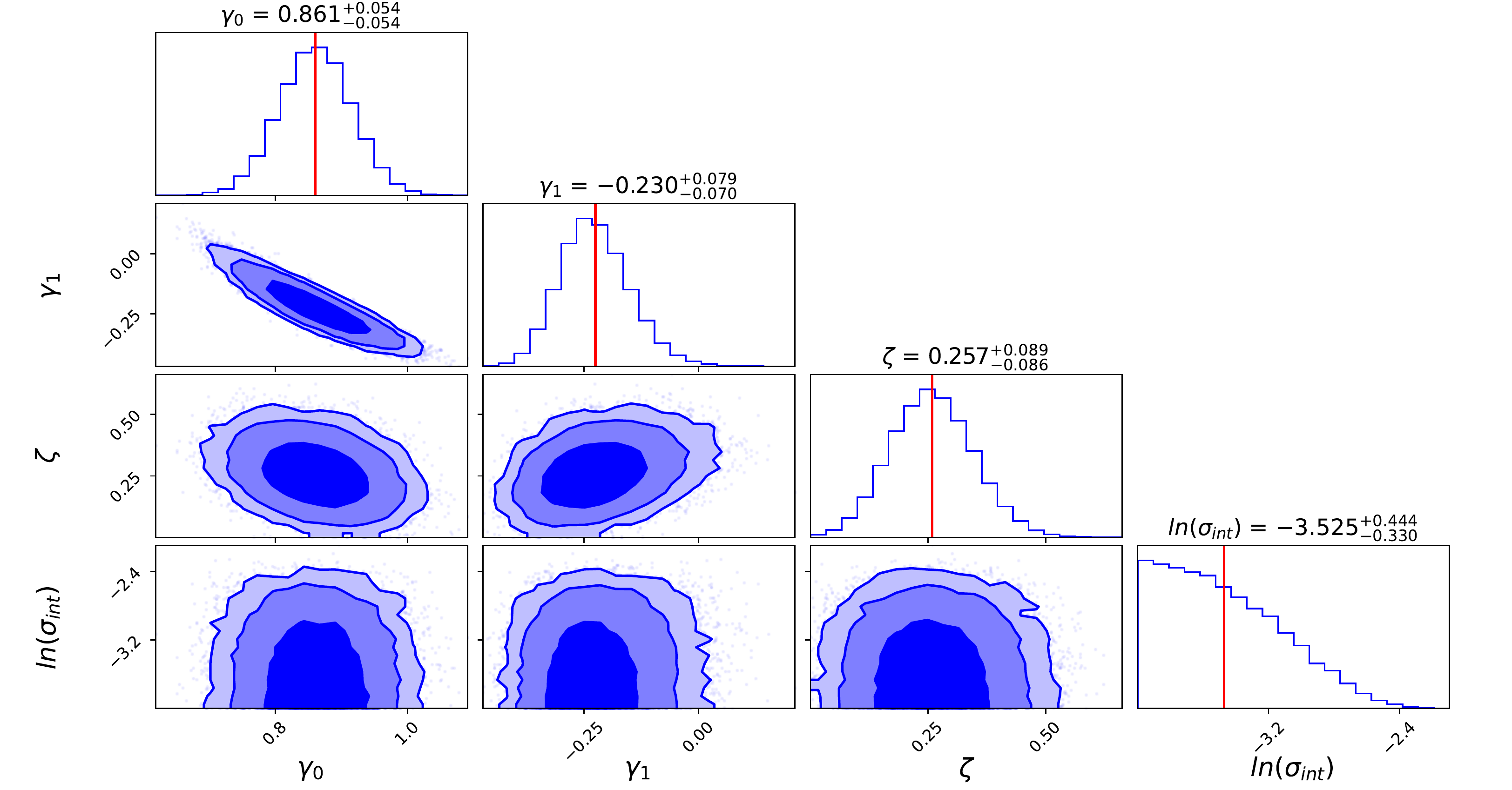}    \caption{Marginalized credible intervals (68\%, 95\%, 99\%) for $\gamma_0$, $\gamma_1$, $\zeta$, and $\ln (\sigma_{int})$ for the SPT-SZ dataset. The parameter $\zeta$ encodes the variation of the gas depletion factor as a function of $M_{500}$ (cf. Eq.~\ref{eq:gammavariationwithmass}).  }
 \label{fig:f14}
\end{figure*}

\begin{figure*}[h]
    %\centering
 \includegraphics[width=12cm, height=10cm]{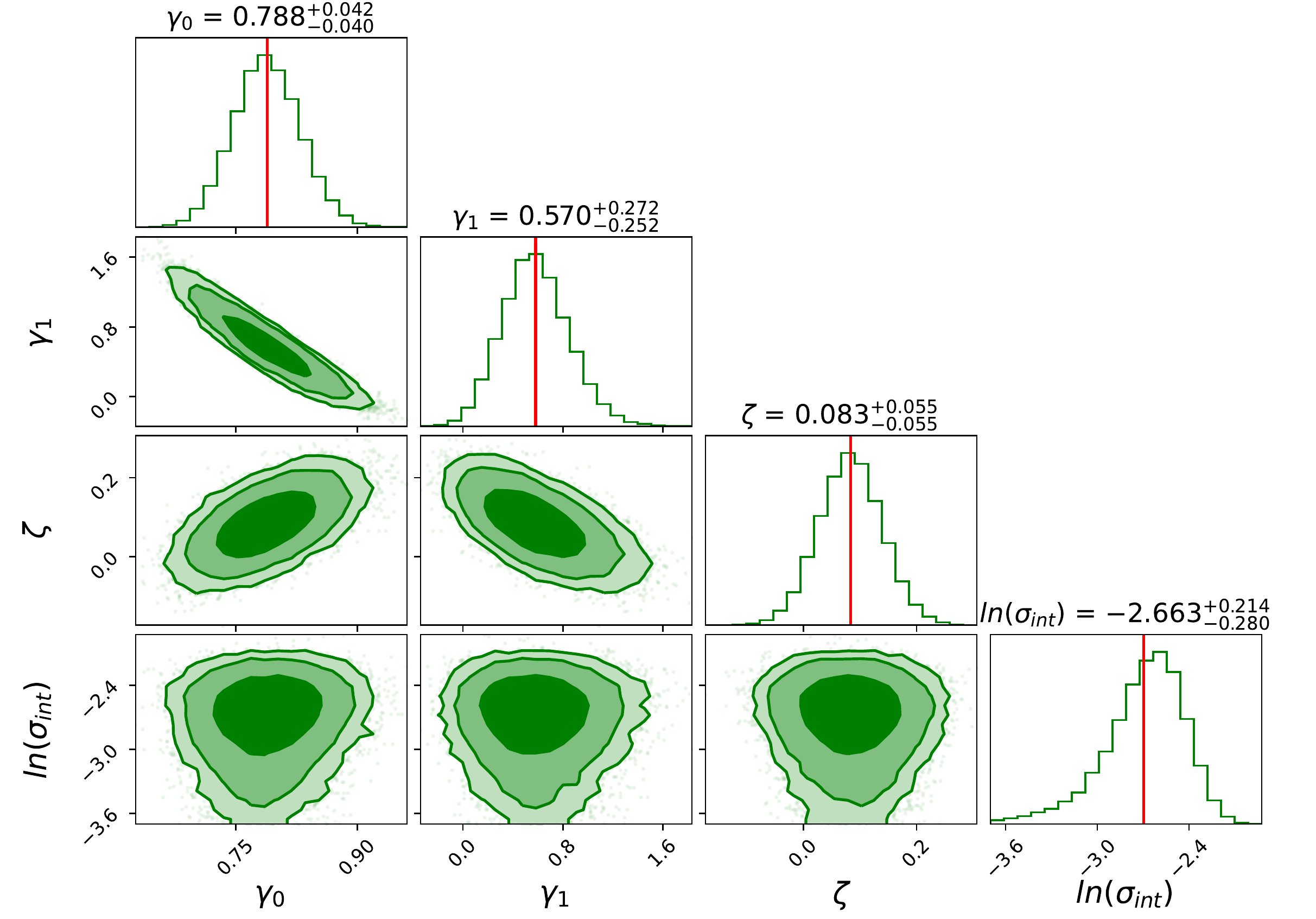}    \caption{ Same as Fig.~\ref{fig:f14} for the Planck ESZ dataset. }
 \label{fig:f15}
\end{figure*}

 Given the contradictory results between the two datasets, we carried out a series of cross-checks on our results using various extended tests, as outlined in Sect.~\ref{sec:dicussions}. We used simple parametric forms to model only the gas mass fraction, without any explicit dependence on the angular diameter distance (cf. Eqs.~\ref{eq:9},~\ref{eq:10},~\ref{eq:11}.)
 We then evaluated the gas depletion factor for both the datasets within the overlapping redshift range. Finally, we also added an explicit dependence on $M_{500}$ in the parameterization used for the gas depletion factor (cf. Eq.~\ref{eq:gammavariationwithmass}). The results of these extended tests are tabulated in Table~\ref{tab:extendedanalysis} and in Fig.~\ref{fig:f14} (SPT-SZ) and Fig.~\ref{fig:f15} (Planck ESZ). We find that the coefficients for the redshift dependent terms in Eqs.~\ref{eq:9},~\ref{eq:10},~\ref{eq:11} are consistent with no variation within 1$\sigma$. For Planck-ESZ, $f_1$ in Eq.~\ref{eq:9} is 5$\sigma$ deviant with respect to zero, which is one of the main underlying causes for a positive value for $\gamma_1$. If we analyze the evolution of the gas depletion factor for a subset of SPT-SZ and Planck ESZ clusters within the same redshift range, we find that $\gamma_1$ is consistent for both the datasets within $1\sigma$ (cf. last two rows in  Table~\ref{tab:summary_table}). Finally, when we allow for  a variation of the gas depletion factor with halo mass, we find that the decreasing/increasing trends with redshift for SPT-SZ (Planck ESZ) still persist, albeit with reduced significance of about $2-3\sigma$.
 
 \rthis{Therefore, our investigations on a model-independent test of the evolution of $f_{gas}$ reveal that the redshift evolution of the gas depletion factor depends on the redshift range,  and also changes depending on whether one includes/excludes a dependence on the halo mass. One possible reason for these changes is that the sample analyzed here is not the full SPT-SZ or  the Planck SZ sample as the optical/X-ray followups  is still ongoing.  Another possibility is that  the error in  $f_{gas}$ also depends on additional systematics such as observable to halo mass relation (which in turn is somewhat cosmology dependent), gas temperature and density profiles, hydrostatic equilibrium assumption, magnetic fields, thermal evaporation, which have not been completely accounted for, and could be redshift dependent. 
 Nevertheless, given the observed  deviation from self-similarity found for both the cluster samples (and in C18)  and the significant change in the redshift dependence of $f_{gas}$ without reference to any underlying cosmology, one must include the dependence in halo mass for modelling  any evolution of $f_{gas}$ with redshift.}
 
 We note that, our results do not affect any cosmological parameter estimations with $f_{gas}$,  since most of those results were obtained with $f_{gas}$  estimated at $R_{2500}$~\cite{allen02,allen04,allen08,allen11}, whereas we used $f_{gas}$ measurements  at $R_{500}$.  Only a handful of works have used gas fraction measurements at $R_{500}$ for cosmology studies~\cite{Pen,ettori09}.
 However, if  the conflicting results for the two SZ datasets at $R_{500}$ \rthis{persist when extended to the full SZ dataset from both the telescopes, then it implies that  one cannot use  $f_{gas}$ values  at $R_{500}$ as a stand-alone probe  for any cosmological tests.}

\section*{ACKNOWLEDGEMENT}
We are grateful to Xiaogang Zheng for useful correspondence about Z19 and to the anonymous referee for many constructive suggestions on the manuscript, which helped us improve the draft.
KB would like to acknowledge the Department of Science and Technology, Government of India for providing the financial support under DST-INSPIRE Fellowship program.

\bibliography{ref}

\iffalse

\fi
 
\end{document}